\documentclass[twocolumn,showpacs,preprintnumbers,amsmath,prl,superscriptaddress]{revtex4}

\usepackage{amssymb}

\usepackage{graphicx}

\def\gz{\ifmmode{Z\hskip -4.8pt Z} 
    \else{\hbox{$Z\hskip -4.8pt Z$}}\fi}
 
\newcommand{\be}{\begin{equation}}
\newcommand{\ee}{\end{equation}}
\newcommand{\bea}{\begin{eqnarray}} 
\newcommand{\eea}{\end{eqnarray}}

\begin{document}

\title{Detection of topological transitions by transport through
molecules and nanodevices}

\author{A.~A. Aligia$^1$, K. Hallberg}
\affiliation{Centro At\'omico Bariloche and Instituto Balseiro,
Comisi\'on Nacional de Energ\'ia At\'omica, 8400 Bariloche, Argentina}
   
\author{B. Normand}
\affiliation{D\'epartement de Physique, Universit\'e de Fribourg,
CH-1700 Fribourg, Switzerland}
  
\author{A.~P. Kampf}
\affiliation{Institut f\"ur Physik, Theoretische Physik III,
Elektronische Korrelationen und Magnetismus, \\ Universit\"at Augsburg,
86135 Augsburg, Germany}

\begin{abstract}

We analyze the phase transitions of an interacting electronic system 
weakly coupled to free-electron leads by considering its zero-bias 
conductance. This is expressed in terms of two effective impurity models 
for the cases with and without spin degeneracy. We demonstrate using 
the half-filled ionic Hubbard ring that the weight of the first 
conductance peak as a function of external flux or of the difference 
in gate voltages between even and odd sites allows one to identify 
the topological charge transition between a correlated insulator and 
a band insulator.

\end{abstract}

\pacs{73.23.-b, 81.07.Nb, 81.07.Ta, 73.40.Qv}

\maketitle


Progress in nanotechnology has made it possible to perform transport
experiments on systems as small as single molecules \cite{molexp}.
Metallic \cite{coll} or semiconducting \cite{aliv,kouw,kou2} quantum 
dots (QDs) can now be assembled into artificial molecules \cite{kouw} 
or solids \cite{coll}. QD molecules of different materials and sizes 
are now being investigated and a wide range of new QD systems is 
expected to be synthesized in the near future \cite{aliv,kouw}. 
The transport properties of a finite chain of 15 QDs were first 
measured over 10 years ago \cite{kou2}, and the metal-insulator 
transition has been studied experimentally in a hexagonal lattice 
of Ag QDs \cite{coll}. These advances open the route for new 
approaches to investigate novel phenomena and theoretical concepts in 
interacting electron systems.

Here we focus on one such phenomenon, the topological phase transition 
associated with a parity change of the ground state. The ionic Hubbard 
model (IHM) with alternating diagonal energy $\pm {\textstyle \frac{1}{2}} 
\Delta$ has received much recent attention \cite{res,fab,tor,kampf,man}. 
It was proposed to describe the neutral-ionic transition in organic 
charge-transfer salts \cite{hubnag}, and later applied to model 
ferroelectricity in perovskites \cite{res,egam}. At half filling and in 
the atomic limit ($t \rightarrow 0$), the ground state is a band insulator 
(BI) for $U < \Delta$, but is a Mott insulator (MI) for $U > \Delta$. 
In one dimension, with non-zero hopping $t$, a spontaneously dimerized 
insulator (SDI) phase appears between BI and MI. With increasing $U$, one 
finds first a charge transition at $U = U_{c}$ from BI to SDI, followed by 
the closing of the spin gap at $U_{s} > U_{c}$, where the transition to the 
MI occurs. Although in finite systems conventional order parameters such as 
charge and spin structure factors vary continuously at a phase transition, 
in a system of $L$ sites one may define charge and spin topological numbers 
which change discontinuously at $U_{c}(L)$ and $U_{s}(L)$ \cite{tor}. 
The charge Berry phase has a step at $U_{c}(L)$, where a 
parity change of the ground state occurs for periodic (antiperiodic) 
boundary conditions if $L = 4m$ ($L = 4m+2$, $m$ integer) \cite{res,tor}.

Here we show that this charge transition can be detected using the total 
intensity or width $w$ of the first peak in a zero-bias conductance 
measurement performed on a flux-threaded ring. The importance of the ring 
geometry is that the topological transition is absent in an open system 
\cite{kampf}. In one of the two parity sectors, which is selected using 
an applied gate voltage, $w \rightarrow 0$ for applied flux $\phi 
\rightarrow 0$ if $L = 4m$ ($\phi \rightarrow \phi_0/2$, where $\phi_0 
= hc/e$ is the flux quantum, if $L = 4m + 2$). Thus the transition, which 
is observed by varying parameters such as $\Delta$, may also be studied 
in ring-shaped molecules where it is not possible to attain a significant 
threading flux.

\begin{figure}[t!] 
\centerline{\includegraphics[height=2.2cm]{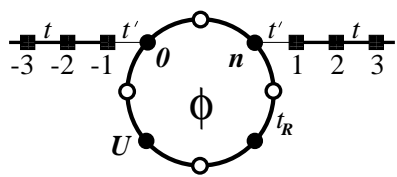}} 
\caption{Interacting electron system on a flux-threaded ring, connected 
by weak links ($t^{\prime }$) to two conducting leads.} 
\end{figure} 

The configuration for the proposed experiment is illustrated in Fig.~1. 
The interacting electron system on the ring is connected to conducting 
leads through two sites, labeled 0 and $n$, by weak links with hopping 
$t^{\prime}$, and either two materials or two different gate voltages 
at alternating sites may be used to model the IHM. It is essential to
distinguish between two cases depending on the spin degeneracy of
the isolated interacting system for odd particle number. We thus 
perform two different calculations of the conductance. The first 
assumes that spin degeneracy is lifted by a Zeeman interaction, but 
allows orbital degeneracy of the states, which is important when including 
interference effects arising in a ring geometry \cite{jagla}. In the 
second, where spin degeneracy is retained, we map the problem into an 
effective Anderson model (EAM) and express the conductance as a function 
of the spectral density of the latter. 

If the ground state $|g \rangle$ of the interacting system is nondegenerate, 
for small $t'$ the links can be eliminated by a canonical transformation,
leading to an effective, non-interacting Hamiltonian for the two leads in 
which the two ``impurity'' sites $i$ connected to the links ($-1$ and 1 in 
Fig.~1) have an energy shift $\Delta \epsilon_{i} (\omega)$, and are 
connected by an effective hopping $t_{\mathrm{eff}} (\omega)$, 
\begin{eqnarray}
H_{\mathrm{eff}} & = & \sum_{k_{L},\sigma }\epsilon _{k_{L}}c_{k_{L} 
\sigma}^{\dag }c_{k_{L}\sigma }+\Delta \epsilon _{-1}c_{-1\sigma }^{\dag} 
c_{-1\sigma }+\Delta \epsilon _{1}c_{1\sigma }^{\dag }c_{1\sigma } \nonumber 
\\ & + & \sum_{k_{R},\sigma }\epsilon _{k_{R}}c_{k_{R}\sigma }^{\dag} 
c_{k_{R}\sigma }+\sum_{\sigma }(t_{\mathrm{eff}}c_{1\sigma }^{\dag} 
c_{-1\sigma }+\mathrm{H.c.}).  \label{heff}
\end{eqnarray}
$L$ ($R$) refers to the lead containing the site $-1$ (+1), and the 
impurity parameters may be expressed in terms of the Green functions for 
the isolated ring $g_{ij}(\omega) = \langle \langle c_{i\sigma }; 
c_{j\sigma }^{\dag} \rangle \rangle$, 
\begin{eqnarray}
\Delta \epsilon_{-1}(\omega) & = & t^{\prime 2} g_{00}(\omega),\text{ } 
\Delta \epsilon_{1}(\omega) = t^{\prime 2} g_{nn} (\omega),  \nonumber \\
&&\text{ }t_{\mathrm{eff}}(\omega) = t^{\prime 2} g_{n0} (\omega). \label{par}
\end{eqnarray}

The transmittance of $H_{\mathrm{eff}}$ may be calculated in different ways.
Without affecting the essential results, we assume identical leads and, 
for their states without sites $(-1,1)$, we consider two models for the 
density of states $\rho (\omega)$ and the hybridization $V_j (\omega)$ 
with sites $(-1,1)$. I) $\rho (\omega)$ and $V_{j}(\omega)$ constant: for 
$j = \pm 1$ and $t_{\mathrm{eff}} = 0$, $g_{jj}^{\mathrm{eff0}}(\omega) = 
1/(\omega - \Delta \epsilon_{j} + i \Gamma)$ with $\Gamma = \pi \rho V^{2}$. 
II) Leads described by a tight-binding model with nearest-neighbor hopping 
$t$, where $g_{jj}^{\mathrm{eff0}} (\omega) = 1/(\omega/2 - \Delta 
\epsilon_{j} + i \Gamma (\omega))$ with $\Gamma (\omega) = \pi \rho 
(\omega) V^{2} (\omega) = \sqrt{t^{2} - \omega^{2}/4}$. By introducing an 
integer $m = 1$ or $2$ for cases I and II,
\begin{eqnarray}
& & T(\omega, V_{g}) = \label{ecf} \\
& & \frac{4 \Gamma^{2}(\omega) |t_{\mathrm{eff}}|^{2}}{|(\frac{\omega}{m}
- \Delta \epsilon_{-1} + i\Gamma (\omega))(\frac{\omega}{m} - \Delta 
\epsilon_{1} + i\Gamma (\omega )) - |t_{\mathrm{eff}}|^{2}|^{2}}. \nonumber
\end{eqnarray}
$V_{g}$ is a gate voltage which changes the on-site energy of all sites 
of the ring by $-eV_{g}$. In case II this equation is exact ($\forall 
\,t^{\prime }$) in the non-interacting system, and generalizes a previous 
result \cite{jagla}. Eq.~(\ref{ecf}) also generalizes previous approaches 
in which only one intermediate state of the ring is included \cite{chen,staf}, 
and is appropriate for the study of interference effects \cite{jagla}. Spin 
degeneracy limits its validity to magnetic fields $B$ for which the Kondo 
effect is destroyed, as discussed below. For sufficiently large Zeeman 
energy $g\mu_{B} B$, the zero-bias conductance at steady state is given by 
$G = G_{0} T(\mu, V_{g})/2$, where $G_{0} = 2e^{2}/h$ and $\mu$ is the 
chemical potential of the leads. The results of Ref.~\cite{izum} suggest 
that Eq.~(\ref{ecf}) remains valid, with $G = G_{0} T(\mu, V_{g})$, also 
in the absence of Zeeman splitting in an intermediate temperature range 
$T_{1} < T < 0.1 w$, where $T_{1}$ is very small. Henceforth we set 
$\mu = 0$, corresponding to half-filled leads. 

The transmittance as a function of gate voltage is very small except at the 
poles of $g_{n0}$. If the ground state $|g \rangle$ has $N$ particles for 
$\mu = 0$, the ground-state energy $E_{g}(N)$ satisfies $E_{g}(N) - eV_{g} 
< E_{g}(N-1)$ and $E_{g} (N) + eV_{g} < E_{g}(N+1)$. As $V_{g}$ 
is increased, a pole in $g_{n0} (V_{g})$ is reached, and for larger $V_{g}$ 
the ground state has $N+1$ particles. Similarly, if $V_{g}$ is lowered, a 
pole is reached when $z = eV_{g} + E_{g}(N-1) - E_{g}(N) = 0$. 

Assuming that the state of lowest energy for $N-1$ particles, $|g(N-1) 
\rangle$, is not degenerate, $g_{ij}$ may be expressed as a Laurent 
expansion around $z = 0$, 
\begin{equation}
g_{ij}(z) = \frac{a_{i} \bar{a}_{j}}{z} + \beta_{ij} + \gamma_{ij} z + \dots ,
\label{la}
\end{equation}
where $\beta _{jj}$, $\gamma _{jj}$, $\dots $ are real coefficients and 
\begin{equation}
a_{j} = \langle g(N-1) |c_{j\sigma }| g(N) \rangle .  \label{alpha}
\end{equation}
$\alpha_{j} = |a_{j}|^{2}$ is the spectral weight for a local photoemission
process. Substituting these expressions in Eqs.~(\ref{par}) and (\ref{ecf}), 
retaining terms to lowest nontrivial order in $t^{\prime}/\Gamma (0)$, and 
using that $T(z) \sim O(1)$ for $z \lesssim (t^{\prime})^{2}/\Gamma (0)$, 
yields 
\begin{equation}
T(z) \cong \frac{1}{1 + (w/z)^{2}} \frac{4 \alpha_{0}\alpha_{n}}
{(\alpha_{0} + \alpha_{n})^{2}} + O((t^{\prime }/\Gamma (0))^{2}), \label{tap}
\end{equation}
where the half-width at half maximum peak height is 
\begin{equation}
w = (\alpha_{0} + \alpha_{n})(t^{\prime})^{2}/\Gamma (0).  \label{w}
\end{equation}
Using $G = G_{0} T(\mu,V_{g})$, this expression coincides at sufficiently 
low temperatures with Eq.~(7) of Ref.~\cite{staf}. 

If sites 0 and $n$ are equivalent by symmetry, then $w = 2(t^{\prime})^{2} 
\alpha_{0}/\Gamma (0)$ and the integrated weight 
$I = \int \mathrm{d} z \, T(z) = \pi w$ are both 
proportional to $(t^{\prime })^{2}$ and to $\alpha_{0}$. If $V_{g}$ is
increased instead of decreased, the same result is obtained with 
$c_{j\sigma} \rightarrow c_{j\sigma }^{\dag}$ and $N-1 \rightarrow N+1$ in
Eq.~(\ref{alpha}). Thus a single transport measurement gives simultaneous
spectral information related to photoemission and to inverse photoemission.
We stress that this information is obtained with much finer energetic 
resolution ($\mu$eV) than that available by direct spectroscopic techniques 
(meV). 

\begin{figure}[t!] 
\centerline{\includegraphics[height=4.0cm]{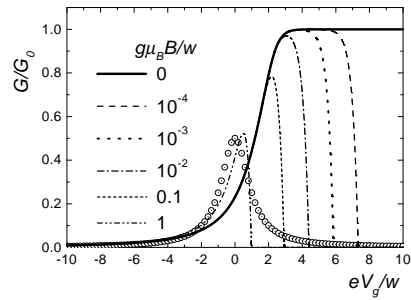}} 
\caption{Conductance as a function of gate voltage for the EAM 
at several magnetic fields. Circles correspond to the 
conductance of a resonant level for the majority spin only.}
\end{figure} 

We turn next to a calculation of the conductance for small or zero $B$, 
where either $|g(N)\rangle$ or $|g(N \pm 1) \rangle$ is spin-degenerate.
This degeneracy can be taken into account accurately if $w$ is small 
compared to the separation between groups of spin-degenerate energy 
levels, because the problem becomes equivalent to an EAM. For simplicity 
we take $V_g$ such that a nondegenerate state $|g(N)\rangle$ is close in 
energy to the spin-degenerate state $|g(N - 1) \rangle$, but extension 
to other cases is straightforward. Neglecting other states, the EAM
{\em for any interacting system} between the leads is 
\begin{eqnarray}
H_{A} & = & - t \!\!\! \sum_{\sigma,i\neq 0,1} \!\! c_{i\sigma}^{\dag} 
c_{i-1\sigma} - \sum_{\sigma} d_{\sigma}^{\dagger} \left( t_{-1}^{\prime} 
c_{-1\sigma} + t_{1}^{\prime} c_{1\sigma} \right) + \text{H.c.} \label{ha} 
\nonumber \\ & & + \varepsilon_{d} \sum_{\sigma} n_{d\sigma} + U_{A} 
n_{d\uparrow} n_{d\downarrow} - g\mu_{B} B (n_{d\uparrow} - n_{d\downarrow}),  
\end{eqnarray}
with $n_{d\sigma} = d_{\sigma}^{\dagger} d_{\sigma}$, $t_{-1}^{\prime} = 
t^{\prime} \bar{a}_{0}$, $t_{1}^{\prime} = t^{\prime} \bar{a}_{n}$, 
$\varepsilon_{d} = E_{g}(N) - E_{g}(N-1) - eV_{g}$, and infinite $U_{A}$ 
(justified because $U_{A} = E_{g}(N) + E_{g}(N-2) - 2E_{g}(N-1) 
\gg w$).

The conductance is given by 
\begin{equation}
G = \frac{2e^{2}}{h}\frac{2\pi w|t_{1}^{\prime }t_{-1}^{\prime }|^{2}}{\left[
|t_{1}^{\prime }|^{2}+|t_{-1}^{\prime }|^{2}\right] }\sum_{\sigma } 
\rho _{d\sigma }(\mu),  
\label{ge2}
\end{equation}
where $\rho_{d\sigma}(\omega)$ is the spectral density of the effective 
$d_{\sigma}$ electrons \cite{dots}. The conductance of the EAM
computed as a function of gate voltage for several values of $B$ 
using slave bosons in the mean-field approximation (SBMFA) \cite{kang} is 
shown in Fig.~2. For $B = 0$, $G$ increases abruptly from 0 to $G_{0}$ 
when $z \sim 0$, and remains nearly perfect for higher $V_{g}$ due 
to pinning of the Kondo peak in $\rho_{d\sigma}(\omega)$ at the Fermi level. 
However, because the Kondo energy scale $T_K$ decreases exponentially with 
$V_g$, even for very small $B$ the plateau becomes a broad peak, terminated 
when $T_K < g \mu_B B$, beyond which $G$ falls to zero. The abruptness of 
the fall is an artifact of the SBMFA, but the total width of the feature 
is well described. For larger $B$ the peak width decreases, tending to $w$ 
to recover the previous result: for $g \mu_{B} B \gtrsim w$, 
$G = G_{0} T(z)/2$, with $T(z)$ given by Eq.~(\ref{tap}).

\begin{figure}[t!] 
\centerline{\includegraphics[height=6.0cm]{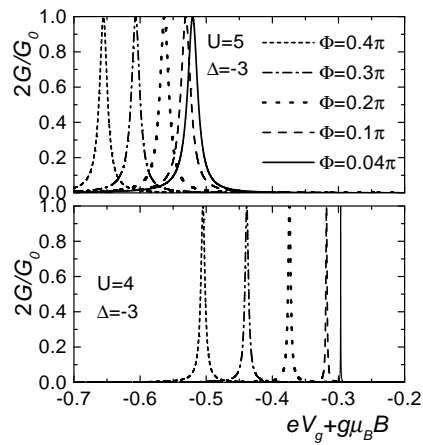}} 
\caption{Transmittance as a function of gate voltage for an 8-site IHM 
ring with leads at sites 0 and 2 for $t_{R} = 1$, $t^{\prime} = 0.2$, 
$\Delta = -3$, and values of $U$ below and above $U_c(L = 8) = 4.352$.}
\end{figure} 

We now apply these results to the topological charge transition in the 
IHM, which is defined by 
\begin{eqnarray}
H_{\rm IHM} & = & - t_{R} \sum_{i,\sigma} (c_{i+1\sigma}^{\dag} c_{i 
\sigma}^{\;} e^{i\Phi /L} + \mathrm{H.c.}) - {\textstyle \frac{1}{2}} \Delta 
\sum_{i} (-1)^{i}n_{i}  \nonumber \\ & & - g \mu_{B} B \sum_{i}(n_{i 
\uparrow} - n_{i\downarrow }) + U \sum_{i} n_{i\uparrow} n_{i\downarrow},  
\label{h}
\end{eqnarray}
where $n_{i\sigma} = c_{i\sigma}^{\dag }c_{i\sigma }^{\;}$, $n_{i} = 
\sum_{\sigma }n_{i\sigma }$, and $N = \sum_{i} n_{i}$. We calculate 
$T(V_{g})$ by exact diagonalization for the isolated ring with $L$ sites 
and $N = L$ electrons, with leads attached to two sites of the same energy 
($-\Delta/2$) in the presence of a magnetic flux $\phi$ (Fig.~1), with 
$\Phi = 2\pi \phi /\phi_{0}$. The Green functions $g_{ij}(V_{g})$ are 
obtained numerically and substituted in Eqs.~(\ref{par}) and (\ref{ecf}). 
We choose $L = 8$, but similar results are obtained for any $L = 4n$ ($L 
= 4n+2$ with $\Phi \rightarrow \Phi + \pi$). The qualitative features are
independent of the lead position $n$, with the exception of $n = L/2$ 
where the transmittance at $\Phi = \pi$ vanishes for symmetry reasons.
Because $H_{\rm IHM}$ is invariant under simultaneous particle-hole 
transformation and sign change of $\Delta $, we restrict our analysis to 
$eV_{g} < 0$. We set $t_{R} = 1$ as the unit of energy unless otherwise 
stated.

The topological transition is present at any value of $t_R$. As a basis 
for our study we consider a realistic ring structure containing 8 QDs with 
only their lowest levels singly occupied, the centers separated by 200 
nm \cite{kou2}, and the parameters $U,\Delta \sim 1$ meV and $t_{R} \sim 
U/4$ (intermediate coupling strength). We first assume the presence of a 
strong magnetic field ($B \sim $1 T for a QD array), in the plane of the 
ring in order not to alter the threading flux, which destroys the Kondo 
effect. Fig.~3 shows the first peak in $G(V_{g})$ at $\Delta = -3$ and 
for two values of $U$. For $L = 8$ and $\Delta = \pm 3$, the topological 
transition occurs at $U_{c} = 4.352$. The peak width is parameter-dependent: 
for $U = 5$, $w$ is approximately constant with decreasing flux, whereas for 
$U = 4$, $w$ decreases and the peak disappears at $\Phi = 0$. If $U < U_{c}$, 
the ground state $|g(L) \rangle$ for $\Phi = 0$ is even under reflection 
through any site (corresponding to a BI), while for $U > U_{c}$ 
$|g(L) \rangle$ is odd (MI or SDI) \cite{tor}. For $\Delta = 0$, the 
lowest-energy hole enters the system with the Fermi wave vector, $\pm 
\pi /2$, leaving an orbitally degenerate ground state with $L-1$ particles, 
$|g(L-1) \rangle $. For $\Delta \neq 0$, this degeneracy is broken and 
$|g(L-1)\rangle $ is odd under reflection through any even site if 
$\Delta$ is negative \cite{note2}. As a consequence, for $U < U_{c}$ 
the matrix element $a_0$ [Eq.~(\ref{alpha})] vanishes by symmetry, 
whence $G$ is negligible at $\Phi = 0$. The flux acts as a symmetry-breaking 
field, which allows one to follow the first peak in a continuous manner 
until it disappears as $\Phi \rightarrow 0$. 
 
\begin{figure}[t!] 
\centerline{\includegraphics[height=4.0cm]{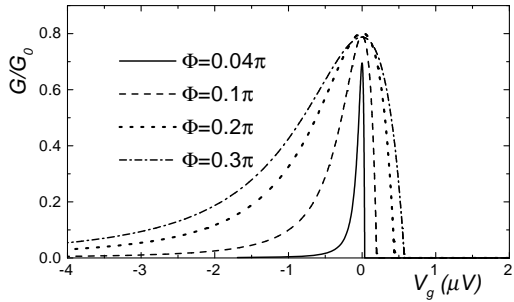}} 
\caption{Transmittance as a function of $V_g$ for a ring of 8 QDs 
described by the IHM. The parameters chosen are $U = 1$ meV, $\Delta = 
-3U/4$, $t=U/4$, $t^{\prime }/t = 0.2$, and $g\mu_{B} = 0.025$ meV/T. The 
curves have been shifted in $V_g$ (cf.~lower panel Fig.~3) such that 
their maxima coincide.}
\end{figure} 

To demonstrate that these essential features are not affected by the 
presence of a Kondo resonance we consider the 8-site ring with parameters 
(Fig.~4) similar to one experimental realization \cite{kou2} and $B$ 
normal to the ring plane so that the threading flux and Zeeman splitting 
have the same origin. Fig.~4 shows the conductance in this regime, 
calculated using the EAM in the SBMFA. The disappearance of the feature 
with decreasing flux remains clear.

The topological transition may now be characterized using $\alpha_0$.
As shown in Fig.~5 for $\Delta = 3$, $\alpha_{0}$ as a function of $U$ 
(cf.~peak widths in Fig.~3 \cite{note2}) vanishes discontinuously at 
$U_{c}$ for $\Phi = 0$, indicating unambiguously the charge transition. 
The analogous result obtained for fixed $U$ by varying $\Delta$ is of 
direct experimental interest, because $\Delta$ can be controlled by a 
difference in gate voltage applied between even and odd sites. A finite 
flux acts as a parity-breaking perturbation and smooths the transition. 
This is the situation for artificial arrays of QDs, in which perfect 
structural symmetry is difficult to attain. We note that curves for all 
flux values cross approximately at the same point: transport measurements 
under different applied fields can therefore help to locate the transition, 
even in the absence of perfect reflection symmetry. However, for a small 
molecule, which by definition exhibits the $\Phi \rightarrow 0$ limit, 
the only source of asymmetry is the leads.

In summary, we have considered the transport properties of a ring-shaped 
interacting system connected weakly to conducting leads. The conductance 
is expressed in terms of the spectral density of an EAM, which illustrates 
the breaking of the Kondo effect by an applied field. Outside the Kondo 
regime we have derived a transmittance formula including nontrivial 
interference effects. The conductance peaks may be characterized by 
their total weight, offering spectral information with far higher 
resolution than conventional spectroscopic measurements. As an 
application of this result, we have demonstrated that the conductance 
can be used to detect the charge transition of molecules or quantum dot 
rings described by the IHM. The method is relevant in the general context 
of systems presenting phase transitions which involve a symmetry change 
of the ground state.

A.A.A. and K.H. are fellows of CONICET, and acknowledge the support of the
Fundaci\'{o}n Antorchas, Project 14116-168, and PICT 03-12742 of ANPCyT.
A.P.K. acknowledges support through SFB484 of the Deutsche
Forschungsgemeinschaft and B.N. the support of the Swiss National Science
Foundation.

\begin{figure}[t] 
\centerline{\includegraphics[height=4.0cm]{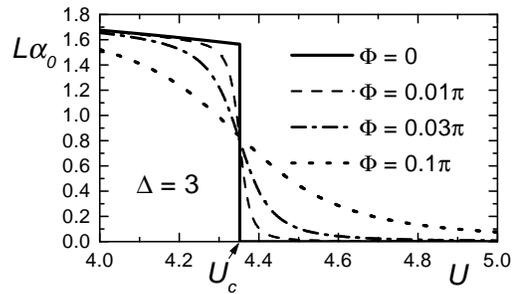}}  
\caption{Matrix element $\alpha_{0}$ as a function of $U$ for $\Delta = 3$ 
($U_c = 4.352$) over a range of values of the flux $\Phi$.} 
\end{figure} 


\begin{thebibliography}{99}
\bibitem{note2}  For reflection through odd sites, $|g(L-1)\rangle $ is odd
for positive $\Delta$. Under a sign change of $\Delta$ the parities of even 
and odd sites are interchanged, and with these the charge sector 
(BI or SDI, MI) in which $\alpha_0 = 0$ at $\Phi = 0$.
\end{thebibliography}
\end{document}